\documentclass[aps,nofootinbib,twocolumn]{revtex4}
\usepackage{graphicx,amsmath,amssymb,amsbsy,latexsym,verbatim}
\usepackage{graphicx}
\usepackage{hyperref}

\newcommand{\be}{\nopagebreak[3]\begin{equation}}
\newcommand{\ee}{\end{equation}}
\newcommand{\ba}{\nopagebreak[3]\begin{eqnarray}}
\newcommand{\ea}{\end{eqnarray}}

\newcommand{\bc}{}

\newcommand{\eq}[1]{(\ref{#1})}

\newcommand{\ket}[1]{\ensuremath{|#1\rangle}}
\newcommand{\bk}[2]{{\langle#1\,|\,#2\rangle}}
\newcommand{\bek}[3]{{\langle#1\,|\,#2\,|\,#3\rangle}}

\newcommand{\C}{\mathbb{C}}
\def\id{1\hspace{-.9mm}\mathrm{l}}

\begin{document}
\title{ \Large A new look at loop quantum gravity}
     \author{Carlo Rovelli}
     \affiliation{Centre de Physique Th\'eorique de Luminy\footnote{Unit\'e mixte de recherche (UMR 6207) du CNRS et des Universit\'es de Provence (Aix-Marseille I), de la M\'editerran\'ee (Aix-Marseille II) et du Sud (Toulon-Var); laboratoire affili\'e \`a la FRUMAM (FR 2291).}, Case 907, F-13288 Marseille, EU}
\date{\small  \today}
\begin{abstract}

\noindent
I describe a possible perspective on the current state of loop quantum gravity, at the light of the developments of the last years.  I point out that a theory is now available, having a well-defined background-independent kinematics and a dynamics allowing transition amplitudes to be computed explicitly in different regimes.  I emphasize the fact that the dynamics can be given in terms of a simple vertex function, largely determined by locality, diffeomorphism invariance and local Lorentz invariance. I emphasize the importance of approximations. I list open problems. 
\end{abstract}

\maketitle

\section{Introduction}
Significant developments in the last years have modified the state of the art in quantum gravity.  The merge of the canonical and the covariant frameworks has yielded a rather well-developed background-independent theory, with a reasonable kinematics and an intriguing dynamics, where physical transition amplitudes can be explicitly computed and compared with the classical theory.  Here is an account of the state of this theory, as I understand it today.  

I present the theory without ``deriving it from classical GR" or other ``quantization procedures".\footnote{In my opinion, after many years of attempts to ``quantize general relativity", it is time to leave the ladder behind, and start taking seriously what the various ``quantization procedures" have produced. It is especially so since large overlaps have appeared between the results of the different quantizations techniques (canonical, path integral and others; see Section \ref{derivation}, below). I expect that it is now going to be more productive to study the theory and its consequences, in order to asses its viability, rather than keep trying to ``derive" the theory.}   As emphasized by Vincent Rivasseau \cite{Rivasseau:2008}, a formulation of quantum field theory that remains meaningful in the background-independent context, is as a generating function for amplitudes associated to a combinatorial structure, as in the definition of QED in terms of Feynman-graphs. The amplitudes define the dynamics by assigning  probabilities to processes described in terms of a Hilbert space.  I use this language here. 

I emphasize in particular the fact --pointed out by Eugenio Bianchi \cite{Bianchi:2010Nice}-- that the dynamics of the theory has a very simple and natural definition, largely determined by general physical principles. It is given by a natural immersion of $SU(2)$ representations into $SL(2,\C)$ ones. A simple group theoretical construction (Eq.~(\ref{vertex}) below) appears to code the full Einstein equations.%
\footnote{{\em Note added in proofs:} For a much simpler and straightforward presentation of the dynamics of the theory, which does not require the full intertwiner space machinery, see \cite{Rovelli:2010vv}.}

I mention below some possible alternatives in the definition of the theory. These are written in {\footnotesize  smaller characters.}\footnote{I do not view alternatives as problems, I view them as opportunities.  In quantum gravity we are not in the embarrassment of riches: we \emph{do not} have numerous complete and consistent theories. In fact, we haven't any.  The theory described here, too, in spite of the various results it yields, is incomplete: a list of open problems is in Section \ref{problems}. At the present state of our knowledge, worries about under-determinacy of the theory are, in my opinion, ill-judged.  Rather than worrying whether this theory might have alternatives, or continuing to sketch new very incomplete models, we better ask if we have at least one complete consistent theory.  This is hard enough, and, in my opinion, is today the relevant  scientific question, and the one likely to be fruitful.} \\

{\footnotesize I take responsibility for the presentation, but the results reported below are due to a number of people, including:  Emanuele Alesci, Abhay Ashtekar, John Barrett,  Eugenio Bianchi, Florian Conrady, You Ding, Bianca Dittrich, Richard Dowdall, Jonathan Engle, Winston Fairbairn, Cecilia Flori, Laurent Freidel, Kristina Giesel, Henrique Gomes, Frank Hellmann, Wojciech Kaminski, Marcin Kisielowski, Kirill Krasnov, Etera Livine, Jurek Lewandowski, Elena Magliaro, Leonardo Modesto, Daniele Oriti, Roberto Pereira, Alejandro Perez, Claudio Perini, Lee Smolin, Simone Speziale, Thomas Thiemann, and Francesca Vidotto.  

I am particularly indebted with Daniele Oriti for a sharp critical reading of these notes and numerous inputs. 

}

\section{Hilbert space and operators}\label{Hs}

The kinematics of a quantum theory is given by a Hilbert space carrying an algebra of operators that have a physical interpretation in terms of observables quantities of the system considered.  These are defined in this section. 

\subsection{Hilbert space}

The Hilbert space  ${\cal H}$  on which the theory is defined is the direct sum of ``graph spaces"
\be
    \tilde{\cal H}= \bigoplus_\Gamma\ {\cal H}_\Gamma
    \label{graphspaces}
\ee\\[-3mm]
factored by an equivalence relation  ${\cal H}=\tilde{\cal H}/\!\!\!\sim$.
The sum \eq{graphspaces} runs over the abstract graphs $\Gamma$. An abstract graph $\Gamma$ is defined by a set of $L$ links $l$, a set of $N$ nodes $n$, together with two functions assigning a source node $s(l)$ and a target node $t(l)$ to every link $l$. 
The graph Hilbert space $ {\cal H}_\Gamma$ is defined to be 
\be
 {\cal H}_\Gamma= L_2[SU(2)^L/SU(2)^N]
\label{n-1}
\ee
where the $L_2$ measure is the Haar measure and the action of $SU(2)^N$ on the states $\psi(U_l)\in L_2[SU(2)^L]:=\tilde{\cal H}_\Gamma$ is 
\be
\psi(U_l) \to \psi(V_{s(l)}U_l V_{t(l)}^{-1}),\hspace{2em}ÊV_n\in SU(2)^N.
\label{gauge}
\ee
These are the `local $SU(2)$ gauge transformations' of the theory.

${\cal H}$  is obtained by factoring $\tilde{\cal H}$ by the equivalence relation $\sim$,  defined as follows.  If $\Gamma$ is a subgraph of $\Gamma'$ then ${\cal H}_\Gamma$ can be naturally identified with a subspace of  ${\cal H}_{\Gamma'}$. Two states are equivalent if they can be related (possibly indirectly) by this identification, or if they are mapped into each other by the discrete group of the automorphisms of $\Gamma$  (maps from links to links and from nodes to nodes that preserve the source and target relations).

This completes the construction of the Hilbert space of the theory. \\[-2.mm]

{\footnotesize \emph{Comments.}  This is the ``combinatorial $\cal H$". An alternative studied in the literature is to consider embedded graphs in a fixed three-manifold $\Sigma$ --namely collections of lines $l$ embedded in $\Sigma$ that meet only at their end points $n$-- and to define $\Gamma$ as an equivalence class of such embedded graphs under diffeomorphisms of $\Sigma$. This choice defines the ``Diff $\cal H$". A further alternative is to do the same but using \emph{extended} diffeomorphisms \cite{Fairbairn:2004qe}. This choice defines the ``Extended Diff $\cal H$". With these definitions a graph is characterized also by its knotting and linking. (If $\Sigma$ is chosen with non-trivial topology, also by the homotopy class of the graph). In addition, with the first of these alternatives graphs are characterized by moduli parameters at the nodes as well (extended diffeos factor away these moduli  \cite{Fairbairn:2004qe}). Neither knotting or linking, nor these moduli, have found a physical meaning so far, hence I tentatively prefer the combinatorial definition.  

The space Diff $\cal H$ is non-separable, leading to a number of complications in the construction of the theory.  The combinatorial $\cal H$ considered here and the extended-Diff $\cal H$ are separable.

Another option is to restrict the theory to graphs $\Gamma$ where all nodes are four valent. (The valence of a node $n$ is the number of links for which $n$ is the source plus the number of links for which it is the target.)  I do not take this option here, although several of the results in the literature refer to the theory restricted in this manner.

}

\subsection{Gravitational field, area, volume and holonomy operators}

The gravitational field $g_{\mu\nu}$ has the dimensions of an area.\footnote{
This follows from $ds^2=g_{\mu\nu}dx^\mu dx^\nu$ and the fact that it is rather unreasonable to assign dimensions to the coordinates of a general covariant theory: coordinates are functions on spacetime, that can be arbitrarily nonlinearly transformed.}
The dimension of the Ashtekar's electric field $E$ (the densitized inverse triad), is also an area. It is convenient to fix units where the area
\be
             8\pi \gamma\ \hbar G= 1
             \label{units}
\ee
where $\gamma$, the Immirzi-Barbero parameter is a positive real number, $G$ is the Newton constant. 
The following operators are defined on ${\cal H}_\Gamma$.  \\[-2.mm]

{\footnotesize Notice that the operators are defined on the individual spaces ${\cal H}_\Gamma$, not on $\cal H$. This is a departure from textbook quantum theory.  Later I will explain how these operators can nevertheless be used in the same manner as standard quantum operators.

}\vspace{1em}

First, the \emph{gravitational field} operator  $\vec L_l= \{L_l^i\}, i=1,2,3$  is the generator of the left $SU(2)$ action in ${\cal H}_{\Gamma}$.\footnote{$L^i_{\hat l}\psi(U_l)\equiv d\psi(U_l(t))/dt$ where $U_{\hat l}(t)=e^{t\tau_i} U_{\hat l}$ and 
$U_{l}(t)=U_{l}$  $\forall l\ne \hat l$. I use the notation $L_l=L^i_l\tau_i$ where $\tau_i$ is a basis in $su(2)$, say $\tau_i=\frac{i}2 \sigma_i$, where $\sigma_i$ are the Pauli matrices. } 
As will become more clear later, $\vec L_l$ is interpreted as the operator corresponding to the flux of Ashtekar's electric field, or the flux of the inverse triad, across ``an elementary surface cut by the link $l$".  \\[-2.mm]

{\footnotesize   It is convenient to define also ``links with reversed orientation" $l^{-1}$. That is $s(l^{-1})=t(l)$ and $t(l^{-1})=s(l)$. The generator of the right $SU(2)$ action $\vec R_l\equiv U_l \vec L_l U_l^{-1}$ is then associated to the link with reversed orientation: $\vec L_{l^{-1}}=\vec R_{l}$. It follows then immediately from \eq{gauge} that 
\be
\sum_{l\in n} \vec L_{l}=0
\label{gauss}
\ee
where the notation $l\in n$ indicates all \emph{oriented} links $l$ such that $s(l)=n$.

}\vspace{1em}

The \emph{area} operator depends on a ``surface"  cutting the links $l_1,... , l_S$. In the combinatorial context, a ``surface"  $\Sigma$ is a collection of (possibly repeated) links $l$ of $\Gamma$. The area operator is defined as
\be 
      A_\Sigma=   \sum_{l\in\Sigma} \sqrt{L_{l}^iL_{l}^i}.
\ee
Its eigenvalues are (in units \eqref{units})
\be 
      A_\Sigma=   \sum_{l\in\Sigma} \sqrt{j_{l}(j_{l}+1)},
\ee
where $j_l$ are half integers. This expression gives the ``spectrum of the area" of the theory. \\[-2.mm]

{\footnotesize The operator can be generalized to surfaces ``cutting a node". This is not strictly necessary in the combinatorial context, as far as I can see.

}\vspace{1mm}

The \emph{volume} operator depends on a ``region". In the combinatorial context, a ``region" $\!R$ is a collection of nodes $n$. The volume operator is given by 
\be 
      V_{R} =  \sum_{n\in{R}}\ V_n.
      \ee
For a trivalent node $n$, $V_n=0$. For a four-valent node $n$,
\be 
      V^2_n = \frac29\  |\epsilon_{ijk}\  L_{l_a}^i L_{l_b}^j L_{l_c}^k |
\ee
where $l_a, l_b, l_c$ are any three (distinct) of the four links of $n$.\footnote{The factor 
$2/9=2^3/3!^2$ gives the volume of a tetrahedron with faces having areas and normals determined by $L_{l}$; see Section \ref{derivation} below. In \cite{Giesel:2005bk}, 
Kristina Giesel and Thomas Thiemann give an argument for a different factor, corresponding to the volume of a \emph{cube}. I am still confused about this factor. This has no effect on what follows.}  The choice of the triple is irrelevant, as it follows easily from \eq{gauss}.\\[-2.mm]

{\footnotesize As pointed out by Thomas Thiemann and Cecilia Flori \cite{Flori:2008nw}, the definition of the vertex operator for higher valent nodes given in the literature, is unsatisfactory. It is not difficult to define a volume operator for general $n$-valent nodes, which reduces to the one on $(n-1)$-valent nodes when one of the links has zero spin; but this can be done in numerous way, and a fully satisfactory choice is still missing. On this, see \cite{Bianchi:2010gc}. This does not affect what follows.

}\vspace{1em}

Finally, the \emph{holonomy} operator is the multiplicative operator $U_l$ associated to each link $l$. The operators $\vec L_l$ and $U_l$ form a closed algebra. 

\subsection{Spin network basis}

Spin networks states are a convenient basis in $\cal H$. 
The Peter Weyl theorem states that $L_2[SU(2)^L]$ can be decomposed into irreducible representations
\be
 \tilde{\cal H}_\Gamma=L_2[SU(2)^L]= \bigoplus_{j_l}\ \bigotimes_l  ({\cal H}_{j_l}^*\otimes {\cal H}_{j_l}). \label{PW}
\ee
Here ${\cal H}_j$ is the Hilbert space of the spin-$j$ representation of $SU(2)$, namely a $2j+1$ dimensional space, with a basis $|j,m\rangle, m=-j,...,j$ that diagonalizes $L^3$. 
The star indicates the adjoint representation, but since the representations of $SU(2)$ are equivalent to their adjoint, we can forget about the star.\footnote{The star does not regard the Hilbert space itself: it specifies a way it transforms under $SU(2)$.} For each link $l$, the two factors in the r.h.s.~of \eq{PW} are naturally associated to the two nodes $s(l)$ and $t(l)$ that bound $l$, because under (\ref{gauge}) they transform under the action of $V_{s(l)}$ and  $V_{t(l)}$, respectively. 
We can hence rewrite the last equation as 
\be
\tilde{\cal H}_\Gamma= \bigoplus_{j_l}\ \bigotimes_n  \tilde{\cal H}_n
\label{sumrep}
\ee
where the node Hilbert space $\tilde{\cal H}_n$ associated to a node $n$ includes all the irreducible ${\cal H}_j$ that transform with $V_n$ under (\ref{gauge}), that 
is\footnote{More precisely, $\tilde{\cal H}_n = (\bigotimes _{l\in s(n)} {\cal H}^*_l)\otimes (\bigotimes _{l\in t(n)n} {\cal H}_l)$ where $s(t)$ and $t(n)$ are the sets of the links for which $n$ is, respectively, a source or a target.}\\[-6mm]
\be
\tilde{\cal H}_n = \bigotimes _{l\in n} {\cal H}_{j_l}.
\label{calHn}
\ee
The $SU(2)$ invariant part of this space
\be
{\cal H}_n = {\rm Inv}_{SU(2)}[\tilde {\cal H}_{n}].
\label{Hn}
\ee
under the diagonal action of $SU(2)$ is called the ``intertwiner space" of the node $n$. A moment of reflection shows that 
\be
{\cal H}_\Gamma= \bigoplus_{j_l}\ \bigotimes_n {\cal H}_n. 
\ee
Thus, a basis in $\cal H$ is labelled by three sets of ``quantum numbers".  An abstract graph $\Gamma$ up to its automorphisms; a coloring $j_l$ of the links of the graph with irreducible representations of $SU(2)$ different from the trivial one\footnote{Because states with $j=0$ are already included in the Hilbert spaces associated to subgraphs, thanks to the equivalence relation $\sim$.} ($j=1/2, 1, 3/2, 2, ...$); and a coloring of each node of $\Gamma$ with an element $v_n$ in an orthonormal basis\footnote{The operator $V_n$ is well defined on the finite dimensional space ${\cal H}_n$ because it is $SU(2)$ invariant and commutes with the areas. It is convenient to choose a basis in ${\cal H}_n$ that diagonalizes it, and I do so here.} in the intertwiner space ${\cal H}_n$. The states $\ket{\Gamma, j_l, v_n}$ labelled by these quantum numbers are called ``spin network states". 

\subsection{Physical picture} \label{interpr}

Spin network states are eigenstates of the area and volume operators.  A spin network state can be given a simple geometrical interpretation.  It represents a ``granular" space where each node $n$ represents a ``grain" or ``chunk" of space. The volume of each grain $n$ is $v_n$.  Two grains $n$ and $n'$ are adjacent if there is a link $l$ connecting the two, and in this case the area of the elementary surface separating the two grains is $8\pi\gamma \hbar G\, \sqrt{j_{l}(j_{l}+1)}$. \\

\begin{figure}[h]
\centerline{\includegraphics[scale=0.3]{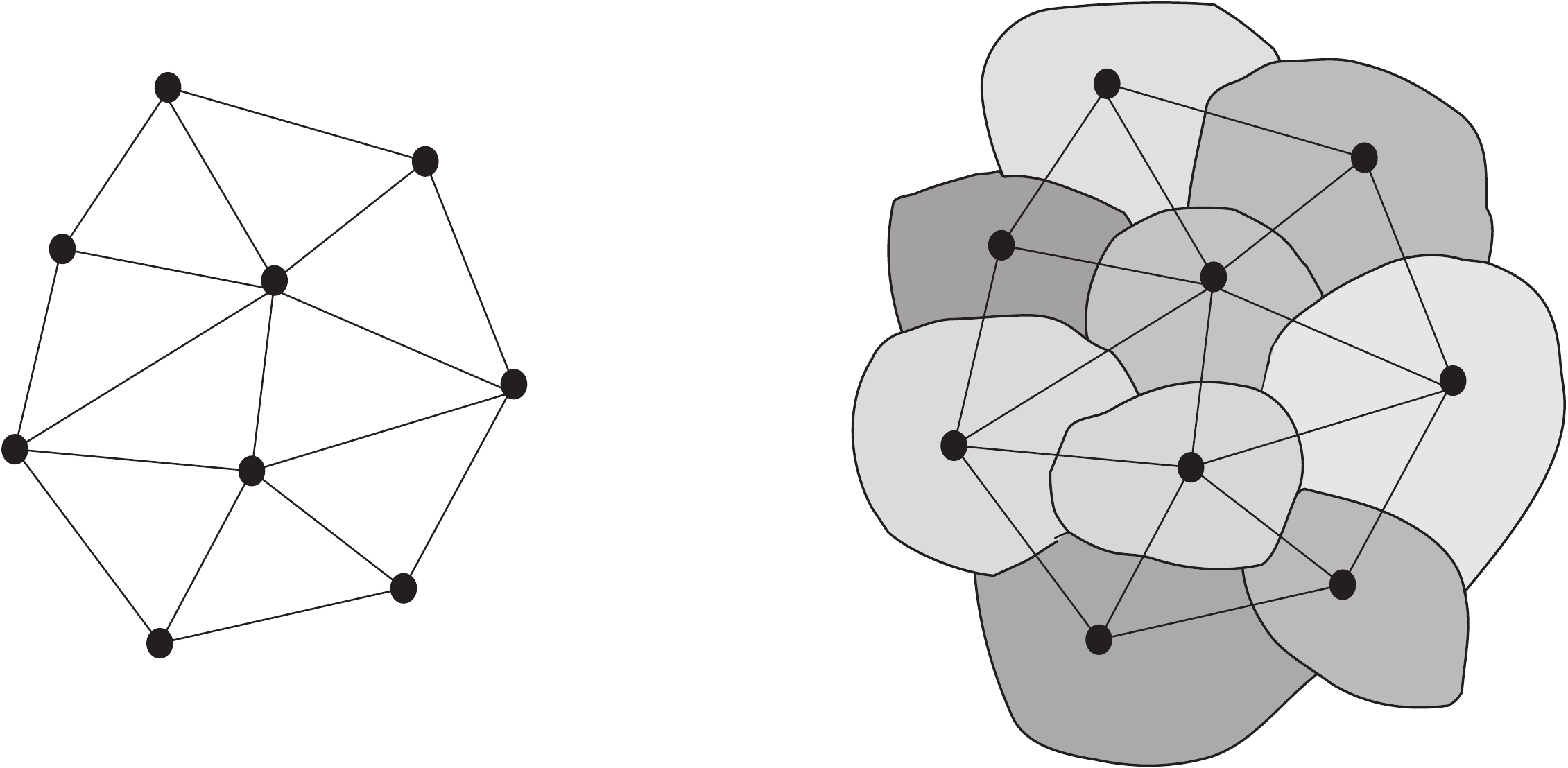}}
$\hspace{-15em}|\Gamma, j_l,v_n\rangle$
\caption{{``Granular" space}. A node $n$ determines a ``grain" or ``chunk" of space.} 
\end{figure}

{\footnotesize  This physical picture admits variants.  In the case $\Gamma$ is the two-skeleton dual to a triangulation of a 3d space, one can view the individual grains as flat tetrahedra.  In some cases, namely for some states, these tetrahedra can be viewed as forming a 3d Regge geometry. (For this, matching conditions between the length of triangles must be satisfied \cite{Freidel:2010aq}.) In the general case, one can associate them a ``twisted geometry" \cite{Freidel:2010aq}. 

Such geometrical pictures are helps for the intuition, but there is no microscopic geometry at the Planck scale and these pictures should not be taken too literally in my opinion. They are choices of classes of continuous geometries interpolating finite sets of geometrical data. It is clear that many such choices are possible. They are analogous to choices of interpolating functions to visualize or describe sets of data points \cite{Rovelli:2010km}.  For instance, we can interpolate a set of data points by means of an interpolating polynomial, or a piecewise linear function, or a piecewise constant functions...  As we will better see below,  these choices have strict analogs for quantum geometry. 

These geometrical pictures can play a very useful role in various situations, but what the theory is about is expectation values of physical observables, not mental pictures of the geometry of individual states.

}\vspace{.7em}

The states in $\cal H$ can be viewed as describing quantum space at some given coordinate time. A more useful interpretation, however, and the one I adopt here, is as describing the quantum space {\em surrounding a given finite 4-dimensional region $\cal R$ of spacetime}.  

This second interpretation is more covariant and will be used below to define the dynamics. That is, states in $\cal H$ are not interpreted as ``states at some time", but rather as ``boundary states".  In the non-general-relativistic limit, therefore, $\cal H$ must be identified with the tensor product ${\cal H}^*_{out} \otimes {\cal H}_{in} $  of the initial and final state spaces of conventional quantum theory. 

Notice that the theory ``knows" about the dimensionality of space only via the fact that the relevant group is $SU(2)$. A remarkable theorem by Penrose, indeed, the spin-geometry theorem \cite{Penrose2}, ensures that the spin-networks constructed from the representations of this group determine a three dimensional geometry.  This theorem is the basis of the construction above.

\subsection{Coherent states and holomorphic representation}

The relation between quantum states and the classical theory is clarified by the construction of coherent states. These are particularly valuable in the present context, where the relation with the classical theory is more indirect than usual. Various classes of coherent states have been studied.  Here I describe the ``holomorphic" coherent states, developed by a number of people  \cite{Hall:2002, Ashtekar:1994nx,Thiemann:2002vj, Bahr:2007xn, Flori:2009rw} and recently discussed in detail by Bianchi-Magliaro-Perini  \cite{Bianchi:2009ky}, as well as the ``semi coherent" states of Livine-Speziale (LS) \cite{Livine:2007vk}.

Holomorphic states are labelled by an element $H_l$  of $SL(2,\C)$ for each link $l$. They are a special case of Thiemann's complexifier coherent states  \cite{Thiemann:2002vj, Bahr:2007xn, Flori:2009rw}. They are defined by 
\be
      \psi_{H_l}(U_l) = \int_{SU(2)^N} dg_n 
       \bigotimes_{l\in \Gamma} K_{t}(g_{s(l)}H_l g^{-1}U_l^{-1}).
\label{states}
\ee
Here $t$ is a positive real number and  $K_t$ is the (analytic continuation to $SL(2,\C)$ of) the heat kernel on $SU(2)$, which can be written explicitly as
\be
K_t(g) = \sum_j (2j+1)  e^{-j(j+1)t}\  {\rm Tr}[D^j(g)]
\ee
where $D^j$ is the (Wigner) representation matrix of the representation $j$. 

The $SL(2,\C)$ labels $H_\ell$ can be given two related interpretations. First, we can decompose each $SL(2,\C)$ label in the form
\be
         H_\ell=e^{2itE_\ell}\ U_\ell
         \label{heu}
\ee 
where $U_\ell\in SU(2)$ and $E_\ell \in su(2)$.  Then it is not hard to show that $U_\ell$  and $E_\ell$ are the expectation values of the operators $U_\ell$ and $L_\ell$ on the state $\psi_{H_\ell}$ 
\be
         \frac{  \bek{\psi_{H_\ell}}{ U_\ell}{\psi_{H_\ell}}  }{  \bk{\psi_{H_\ell}}{\psi_{H_\ell}} }=U_\ell~, \hspace{1em}
         \frac{  \bek{\psi_{H_\ell}}{L_\ell}{\psi_{H_\ell}}  }{  \bk{\psi_{H_\ell}}{\psi_{H_\ell}} }= E^i_\ell~,
\ee 
and the corresponding spread is small.\footnote{
Restoring physical units, $ \Delta U_\ell \sim \sqrt{t}$ and $\Delta E_\ell \sim 8\pi\gamma\hbar G \sqrt{1/t}. $
If we fix a length scale $L\gg \sqrt{\hbar G}$ and choose $t=\hbar G/L^2\ll 1$, we have then $
        \Delta U_\ell \sim ~ \sqrt{\hbar G}/L$ and $
        \Delta E_\ell \sim  \sqrt{\hbar G}\,L$, which shows that both spreads go to zero with $\hbar$.}
        
Alternatively, we can decompose each $SL(2,\C)$ label in the form
\be
H_\ell = n_{s,\ell} ~ e^{-i(\xi_\ell+i\eta_\ell)\frac{\sigma_3}{2}} ~ n^{-1}_{t,\ell}.
\label{fs}
\ee
where $n\in SU(2)$. Let $\vec z=(0,0,1)$ and $\vec n=D^1(n)\vec z$. Freidel and Speziale discuss a compelling geometrical interpretation for the $(\vec n_s, {\vec n}_t, \xi, \eta)$ labels defined on of each link by \eq{fs} \cite{Freidel:2010aq} (see also \cite{Dittrich:2008ar,Oriti:2009wg,Bonzom:2009wm}). For appropriate four-valent states representing a Regge 3-geometry with intrinsic and extrinsic curvature, the vectors ${\vec n}_s, {\vec n}_t$ are the $3d$ normals to the triangles of the tetrahedra bounded by the triangle; $\xi$ is the extrinsic curvature at the triangle and $\eta$ is the area of the triangle divided by $8\pi\gamma G\hbar$. 
  For general states, the interpretation extends to a simple generalization of Regge geometries, that Freidel and Speziale have baptized ``twisted geometries". \\[-1mm]

{\footnotesize Freidel and Speziale give a slightly different definition of coherent states \cite{Freidel:2010aq}. The two definitions converge for large spins, but differ at low spins. It would be good to clarify their respective properties, in view of the possible applications in scattering theory (see below).

}\vspace{1em}

Of great use are also the Livine-Speziale (LS) ``semi-coherent" states. They are defined as follows. The conventional magnetic basis $|j,m\rangle$ with $m=-j,...,j$, in $H_j$ diagonalizes $L^3$. Its highest spin state $|j,j\rangle :=|j,m=j\rangle$ is a semiclassical state peaked around the classical configuration $\vec L=j\vec z$ of the (non commuting) angular momentum operators.  If we rotate this state, we obtain a state peaked around any configuration $\vec L=j \vec n$.  The state 
\be
      |j,n\rangle = D^j(n)|j,j\rangle= \sum_m D^j_{jm}(n)|j,m\rangle,
\label{n}
\ee
is a semiclassical state peaked on $\vec L=j \vec n=j D^1(n)\vec z$.\\[-2.mm]

{\footnotesize
The states \eq{n} are generally denoted as
\be
      |j,\vec n\rangle := |j,n\rangle. 
      \label{vecn}
\ee
where $\vec n =D^j(n)\vec z$. I find this notation confusing. The problem is of course that there are many different $n$ (many rotations) that yield the same $\vec n$, therefore the state $|j,\vec n\rangle$ is not defined by this equation.  The common solution is to choose a ``phase convention" that fixes a preferred rotation $\hat n$ for each $\vec n$. For instance, one may require that $D^1(\hat n)$ leave $\vec z\times \vec n$ invariant. I would find it clearer, even after such a phase convention has been chosen, to still add a label to the notation \eq{vecn}, say for every rotation $n_\phi$ that leaves $\vec n$ invariant, 
\be
      |j,\vec n, \phi \rangle := |j,n_\phi \hat n \rangle =  e^{ij\phi}\ |j,\hat n\rangle.
\ee\\[-1mm]
The reason is that this phase has a physical interpretation: it codes the extrinsic curvature at the face.

}\vspace{.6em}

LS states are states in ${\cal H}_n$, where $n$ is $v$-valent (unfortunate notation: here $n$ indicates a node, not an $SU(2)$ element as above),  labelled by a unit vector $\vec n_l$ for each link $l$ in $n$, defined by  
\be
      |j_l,\vec n_l\rangle = \int_{SU(2)} \!dg\ \bigotimes_{l\in n} D^{j_l}(g)|j_l,\vec n_l\rangle.      
\ee
The integration projects the state on ${\cal H}_n$. These states are not fully coherent:  they are eigenstates of the area, and the observable conjugated to the area (which is related to the extrinsic curvature) is fully spread.

Remarkably, in \cite{Bianchi:2009ky} it is shown that for large $\eta_l$ the holomorphic states are essentially LS states which are also wave packets on the spins. That is
\be
\langle j_l, \vec n_l | \psi_{H_l}\rangle \sim \prod_l e^{-\frac{j_l-j^0_l}{2\sigma_l}}\ e^{i\xi_l j_l}
\ee
where $\vec n$ and $\vec {\tilde n}$ are identified with the $\vec n$ in $s(l)$ and $t(l)$ respectively and where $2j_l+1=\eta_l/t_l$ and $\sigma_l=1/(2t_l)$. Thus, the different coherent states that have been used in the covariant and the canonical literature, and which were long thought to be unrelated, are in fact essentially the same thing.   \\[-1.mm]

In summary, the Hilbert space ${\cal H}_\Gamma$ contains an (over-complete) basis of ``wave packets" $\psi_{H_l}=\psi_{\vec n_l, {\vec n}'_l, \xi_l, \eta_l}$, with a nice interpretation as discrete classical geometries with intrinsic and extrinsic curvature. 

These states define a natural holomorphic representation of ${\cal H}_\Gamma$  \cite{Ashtekar:1994nx,Bianchi:2010mw}. In this representation, states are represented by holomorphic functions on $SL(2,\C)^L$
\be
    \psi(H_l)=\bk{\psi_{H_l}}{\psi}.
\ee

\subsection{Derivation and relation with $SL(2,\C)$}\label{derivation}

Above I have presented the kinematics of the theory without \emph{deriving} it from known physics. Remarkably, there are a number of distinct derivations that converge to this construction. Such convergence provides supports the credibility of this kinematics.  I mention here the main ones of these derivations. This will also allow me to introduce a structure, the map $f_\gamma$, that plays an important role below. 

\begin{enumerate}

\item{\em Canonical quantization.} The strongest reason for taking the kinematical picture described above seriously, in my opinion, is that it is the result of a rather conventional quantization of the phase space of general relativity \cite{Ashtekar:2004eh,Thiemann,Rovelli}.  If we start from the phase space of general relativity in the Ashtekar formulation, choose a Poisson algebra of observables, represent it in terms of operators on a Hilbert space, and factor away the relevant gauge invariances, we obtain  the Hilbert space and the operators constructed above. 

The algebra of observables to choose is formed by holonomies $U_\gamma$ of the Ashtekar connection $A$ around closed loops $\gamma$, and fluxes of the Ashtekar electric field $E$ across surfaces. The Poisson algebra of these operators can be represented by operators acting on a space $\cal S$ of functionals $\psi[A]$ of the connection. The space $\cal S$ is formed by (limits of sums of products of) functionals that depend on the value of $A$ on graphs.

The key gauge invariance is 3d coordinate transformations, which plays three major roles.  First, it is the main hypothesis for a class of theorems stating that the resulting representation is essentially unique \cite{Lewandowski:2005jk,Fleischhack:2006zs}. Second, it ``washes away" the location of the graph $\Gamma$ in $\Sigma$, so that all the Hilbert subspaces associated to distinct but topologically equivalent graphs in $\Sigma$ end up identified.  Depending on the particular class of coordinate transformations one allows in the classical theory, one ends up with the different versions of the Hilbert space mentioned above. Third, this gauge invariance resolves the difficulties that have plagued the previous attempts to use a basis of loop states in continuous gauge theories. 

The other gauge invariance of the canonical theory is formed by the local $SU(2)$ transformations, which gives rise to \eq{gauge}.

\item{\em Polyhedral quantum geometry.}  The idea of polyhedral quantum geometry is to describe ``chunks" of quantum space by quantizing the space $\tilde S$ of the ``shapes" of the geometry of solids figures (tetrahedra, or more general polyhedra) \cite{Barbieri:1997ks,Barrett:1999qw,Barrett:2009cj,Pereira:2010}. This space can be given a rather natural symplectic structure as follows.  Take a flat tetrahedron, for simplicity.   Its shape can be coordinatized by the four normals $\vec L_l, l=1,2,3,4$ to its faces, normalized so that  
 $|\vec L_l|=a_l$ is the area of the face $l$.  A natural $SO(3)$ invariant symplectic structure on $\tilde S$ is $\omega= \sum_l \epsilon_{ijk}\, L_l^i\, dL_l^j\wedge dL_l^k$, or, equivalently, by the Poisson brackets
\be
       \{L_l^i, L_{l'}^j\} =\delta_{ll'}\, \epsilon^{ij}{}_k\  L^k. 
\ee
A quantum representation of this Poisson algebra is precisely defined by the generators of $SU(2)$ on the space $\tilde {\cal H}_n$ given in \eq{calHn} (for a 4-valent node $n$). The operator corresponding to the area  $a_l=|\vec L_l|$ is the Casimir of the representation $j_l$, therefore the space  ``quantizes" the space of the shapes of the tetrahedron with areas $j_l(j_l+1)$. Furthermore, the normals of a tetrahedron satisfy 
\be
       \vec C := \sum\!\raisebox{-1mm}{${}_l$} \ \vec L_l= 0. 
\label{C}
\ee
The Hamiltonian flow of $\vec C$, generates the rotations of the tetrahedron in $R^3$. By imposing equation \eq{C} and factoring out the orbits of this flow, the space $\tilde S$ reduces to a space $S$ which is still symplectic.  In the same manner, imposing the operator equation \eq{C} strongly on $\tilde{\cal H}_n$ gives the space  ${\cal H}_n$ given in \eq{Hn}.    

The construction generalizes to polyhedra with more than 4 faces. Then the shape of an ensemble of such polyhedra, with the same area and opposite normals on the shared faces\footnote{The area and the normals match, but not the rest of the geometry of the face, in general. Thus, we have ``twisted geometries", in the sense of Freidel and Speziale.}, is quantized precisely by the Hilbert space $\cal H$ defined above. 

What is the relation with gravity? The central physical idea of general relativity is of course the identification of gravitational field and metric geometry.  Consider a polyhedron given on a (say, piecewise linear) manifold. A metric geometry is assigned by giving the value of a metric, or a triad field $e^i=e^i_a dx^a$, namely the gravitational field.  Consider the quantity
\be
 E^i_l:=\epsilon_{ijk} \int_l  e^j\wedge e^k.
\ee
Observe that on the one hand this is precisely the flux of the densitized inverse triad $E^{ia}$ across the face $l$ of the polyhedron:
\be
 E^i_l=\int_l \ n_a E^{ai}\,, 
\ee
where $n_a$ is the normal to the face; on the other hand, in locally flat coordinates it is the normalized normal $\vec n_l$ to the face $l$, multiplied by the area:
\be
 E^i_l 
 =\int_l n_a E^{ai}=\int_l n^i = n_l^i a_l = L_l^i.
\ee
Therefore the quantized normals $\vec L_l$ of simplicial quantum geometry can be interpreted as the quantum operator giving the flux of the Ashtekar electric field, and we recover again the full kinematics of the previous section.

\item{\em Covariant lattice quantization.} A third possibility is to discretize general relativity  on 4d lattice with a boundary, and study the resulting Hilbert space of the lattice theory. This is close in spirit to lattice gauge theory. The difference is diffeomorphism invariance: in general relativity the lattice is a ``coordinate" lattice, and coordinates are gauges. Thus for instance there is no analog of the QCD lattice spacing $a$. More precisely, the physical dimensions (lengths, areas, volumes) of the cells of the lattice are not fixed, as in lattice gauge theory, but are determined by the discretized field variables themselves. 

The (double covering of the) local gauge group of the covariant theory is $SL(2,\C)$ and the boundary space that one obtains on the boundary of the lattice theory is
\be
{\cal H}^{SL(2,\C)}_\Gamma= L_2[SL(2,\C)^L/SL(2,\C)^N].
 \label{hsl}
\ee
where $\Gamma$ is the two-skeleton of the boundary of the lattice. 
The states in this Hilbert space $\psi(H_l), H_l\!\in\! SL(2,\C)$, can be seen as wave functions of the holonomies $H_l \!=\! {\cal P}\exp{\int_{l} \omega}$ of the spin connection $\omega$, along the links $l$.  The corresponding generators $J$ of the Lorentz group must therefore represent the conjugate momentum of $\omega$. Since the dynamics of general relativity can be coded into the  Holst action
\be
S=\int [(e\wedge e)^*+\frac1\gamma (e\wedge e)]\wedge F[\omega]
\ee
these momenta are (the projection on the boundary of spacetime of) 
\be
  J=e\wedge e+\frac1\gamma (e\wedge e)^*.
\label{simp3}
\ee
It is easy to show that an $SL(2,\C)$ algebra element $J$ has the form \eq{simp3} iff there is a gauge in which its rotation and boost components\footnote{That is $L^i=\frac12 \epsilon^i{}_{jk}J^{jk}$ and $K^i=J^{0i}$} $\vec L$ and $\vec K$ are related by 
\be
      \vec K = - \gamma \vec L.
              \label{simp}
\ee
This relation is sometime denoted the ``simplicity constraint".\footnote{In general, the  ``simplicity constraints" are the relations $J$ must satisfy in order to have the form \eqref{simp3}. Equation \eqref{simp} is a version of these.} I will return to this important relation shortly. 

\paragraph*{The map $f$.}
The relation between the $SU(2)$ Hilbert space ${\cal H}_\Gamma$ defined in \eq{PW} and the Lorentzian Hilbert space (\ref{hsl}) is important for what follows. There exists a natural immersion of the first into second. To see it, consider again the Peter-Weyl decomposition to write 
\ba
{\ }\ \ \ \ \tilde {\cal H}^{SL(2,\C)}_\Gamma &=&  L_2[SL(2,\C)^L]
\label{PWL}
\\ &=& \sum_{(p_l,k_l)} \bigotimes_{l}  ({\cal H}^*_{(p_l,k_l)}\otimes {\cal H}_{(p_l,k_l)}).\nonumber
\ea
Here $(p\in R, k\in Z^+)$ are the labels of the $SL(2,\C)$ unitary irreducible representations. Now, fix an $SU(2)$ subgroup of $SL(2,\C)$. Then each Lorentz irreducible decomposes into a sum of $SU(2)$ irreducibles 
\be 
{\cal H}_{(p,k)}= \bigoplus_{j'=k}^\infty {\cal H}_{j'}
\ee
The \emph{first} term of this sum ${\cal H}_{j'=k}\subset {\cal H}_{(p,k)}$, namely the lowest-spin irrep, plays a key role below. Consider the map
\be
Y_\gamma: \tilde{\cal H}_\Gamma \to \tilde {\cal H}^{SL(2,\C)}_\Gamma
\ee
defined by sending each $SU(2)$ irreducible of $\tilde {\cal H}_\Gamma$ to the $j'=k$ subspace of the Lorentz irreducible $(p=\gamma j, j)$
\be
   Y_\gamma:  {\cal H}_j \longmapsto {\cal H}_j\subset {\cal H}_{(p=\gamma j,\, k=j)}. 
\label{Y}
\ee
The image of this linear map $Y_\gamma$ is the subspace of $\tilde {\cal H}^{SL(2,\C)}_\Gamma$ obtained by restricting the sum (\ref{PWL}) to the irreducibles where
\be
        p_l=\gamma j_l, \hspace{3em}  k_l=j_l,       
\label{p}
\ee
and restricting each $SL(2,C)$ irreducible to its (finite dimensional) minimum weight subspace $j'_l=k_l=j_l$. 

Now, one can then show by explicit calculation \cite{Ding:2009jq,Ding:2010} that 
\be
        \langle \psi |  \vec K + \gamma \vec L | \phi  \rangle = 0.
\ee
for any $\psi$ and $\phi$ in the image of $Y_\gamma$.
In other words, the image of $Y_\gamma$ is a subspace of $ L_2[SL(2,\C)^L]$ where the constraints \eq{simp} are implemented weakly.  But this is precisely the relation that constrains $J$ to be of the form \eq{simp3}! 

In other words: {\em the image of the natural map \eq{Y} is a subspace where the equation that constrains the momentum $J$ to the form it has in general relativity holds weakly.} 

This image is the correct subspace for defining the quantum theory corresponding to classical GR. One can also verify  \cite{Engle:2007wy,Ding:2009jq,Ding:2010} that the geometrical operators defined in the covariant theory are sent to the corresponding ones of the canonical theory by $Y_\gamma$.

Restricting $Y_\gamma$ to $SU(2)$ invariant states (namely to ${\cal H}_\Gamma$) and composing it with the projection $P_{SL(2,\C)}:\tilde {\cal H}^{SL(2,\C)}_\Gamma\to {\cal H}^{SL(2,\C)}_\Gamma$ on the $SL(2,\C)$ invariant states, defines the map 
\be
     f_\gamma = P_{SL(2,\C)}\circ Y_\gamma :\ \  {\cal H}_\Gamma \to  {\cal H}^{SL(2,\C)}_\Gamma
     \label{f}
\ee
from $SU(2)$ spin networks to $SL(2,\C)$ spin networks. The covariant theory lives on the image of this map. Once again, therefore, we recover the kinematics given in the previous section.\footnote{$f_\gamma$ is injective \cite{Kaminski:2009cc}.} 
\end{enumerate}

This concludes the description of the kinematics of the theory.  It is time to move up to the dynamics.

\section{Transition amplitudes}

In a general covariant quantum theory, the dynamics can be given by associating an amplitude to each boundary state \cite{Oeckl:2003vu,Oeckl:2005bv}. Therefore, the dynamics is given by a linear functional $W$ on $\cal H$. The modulus square 
\be
P(\psi)=|\langle W|\psi\rangle|^2
\ee
is the probability associated to the process defined by the boundary state $\psi$. This is described in detail, for instance, in the book \cite{Rovelli}. 

How is $W$ defined? As pointed out by Eugenio Bianchi in his Nice lectures \cite{Bianchi:2010Nice}, the form of $W$ is largely determined by general principles:  Feynman's superposition principle, locality, diffeomorphism invariance, crossing symmetry, and local Lorentz invariance. Let us discuss these principles and their consequences, one by one.
\begin{enumerate}
\item{\em Superposition principle.} Following Feynman, we expect that the amplitude $\bk{W}\psi$ can be expanded in a sum over ``histories of states" 
\be
\langle W|\psi\rangle=\sum_\sigma \ W(\sigma),
 \label{Wpsi}
\ee
where $W(\sigma)$ is an amplitude associated to an appropriate sequence of states $\sigma$, bounded by $\psi$. Recall that in conventional quantum mechanics expressions like the above one can be derived by inserting resolutions of the identity in the evolution operator, as well as from perturbation theory, as in QED, where scattering amplitudes can be computed by summing amplitudes of processes with a finite number of vertices. 

\item{\em Locality.}
We expect that the amplitude $W(\sigma)$ can be built in terms of products of elementary amplitudes $W_v$ associated to local \emph{elementary} process, as the vertices in QFT 
\be
 W(\sigma)\ \sim \ \prod_v    W_v.
 \label{Wsigma}
\ee
Let us, therefore, focus first on the amplitude $W_v$ of a single \emph{elementary}  process. This will be interpreted as an elementary vertex, in the same sense in which the QED vertex is the elementary dynamical process that gives an amplitude to the boundary Hilbert space of two electrons and one photon. 

\item{\em Diffeomorphism invariance.} I use this expression here in a very loose sense, to denote the following. Recall that in the canonical quantum theory the diffeomorphism invariant dynamics is generated by the hamiltonian constraint. This is a \emph{density}, and therefore, loosely speaking, acts only where there may be volume; that is on the nodes of $\psi$.   Thus $W_v$ must be associated to processes that transform nodes into nodes.\footnote{The fact that the canonical Hamiltonian constraint does not act if there are no nodes is the key result that sparked the interest in loop quantum gravity \cite{Rovelli:1987df,Rovelli:1989za}. In my opinion it is the founding technical result in the loop representation. Because of diffeomorphism invariance, there is nothing physical between node and node. In Einstein's words, if you remove the gravitational field, what remains is not empty space: it is nothing at all.}

Given a spin network state $\ket\psi=\ket{\Gamma,j_l,v_n}$, we can visualize the elementary process that has $\psi$ has boundary state as a single \emph{vertex} (a point), directly connected by \emph{edges} (lines) to the nodes of $\Gamma$ and by \emph{faces} (surfaces) to the links of $\Gamma$.\footnote{The standard terminology is
\emph{nodes} and \emph{links} for the graphs of the spin networks; and 
\emph{vertices}, \emph{edges} and \emph{faces} for the two-complex of the spinfoams.} See figure \ref{vrtx}. 

\begin{figure}[h]
\centerline{\includegraphics[scale=0.3]{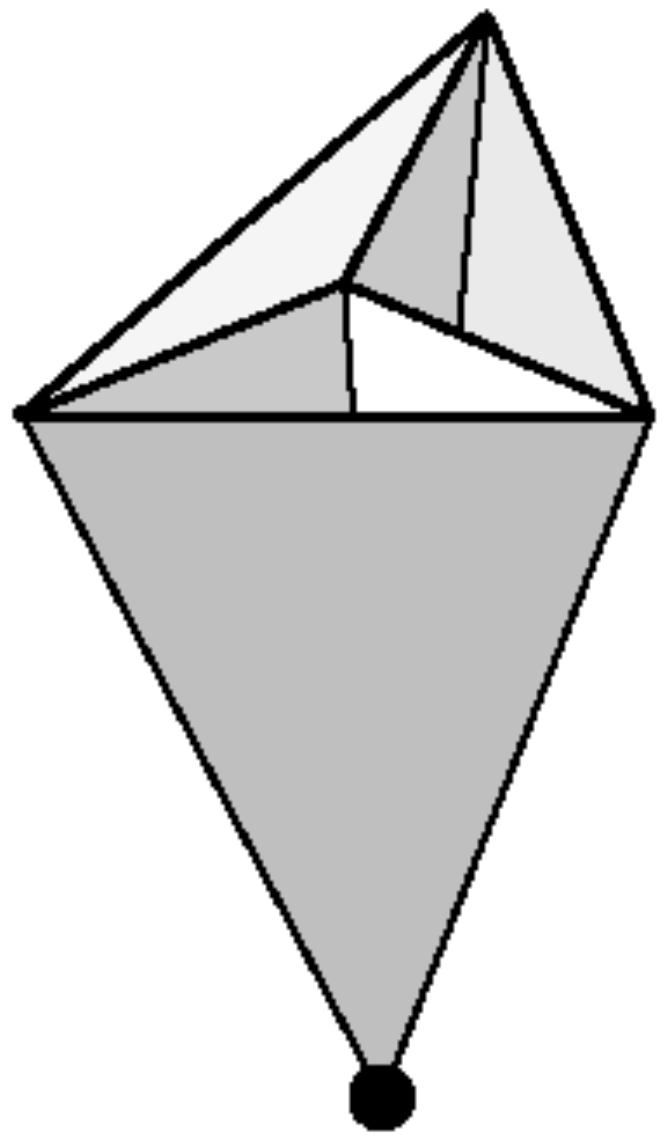}}
\caption{Graphical representation of the elementary vertex, for a boundary spin network with $\Gamma$ formed by the complete graph with 4 nodes (a tetrahedron).}
\label{vrtx}
\end{figure}

The amplitude of this elementary process will be a function $W_v(j_l,v_n)$. This function determines the theory

\item{\em Crossing symmetry.}  It is a well know property of standard QFT that the vertex amplitude does not depend on which states are considered as ``in" and which are considered as ``out".  Assume the same is true in gravity.  

Finally, let me come to the essential ingredient: 

\item{\em Lorentz invariance.} Since classical general relativity has a local Lorentz invariance, we expect the individual spinfoam vertex to be Lorentz invariant in an appropriate sense.  Since the Hilbert space ${\cal H}_\Gamma$ defined above has no hint of $SL(2,\C)$ action, there should be a map from it to a Lorentz covariant language that characterizes the vertex.  How? 

Well, we have just constructed such a map in the previous section: it is the map $f_\gamma$, which depends only on a single parameter $\gamma$. 
\end{enumerate}

I am now ready to define a vertex amplitude that satisfies these requirements. 

\subsection{The LQG vertex}

A simple vertex amplitude that satisfies the above requirements is
\be
   \langle W_v| \psi\rangle =   (f_\gamma\psi)(\id).
\label{vertex}
\ee
Here $f_\gamma$ is the map defined in \eq{Y} and \eq{f}, which takes an $SU(2)$ spin network to an $SL(2,\C)$ spin network. The right hand side is 
the {\em evaluation} of the $SL(2,\C)$ spin network, that is the value $  (f_\gamma\psi)(H_l\!=\id)$,  of the spin network state (in the $\psi(H_l)$ representation) when $H_l$ is equal to the identity for each $l$.

The vertex amplitude \eq{vertex} has been found independently by different research groups \cite{Engle:2007uq,Livine:2007vk,Engle:2007qf,Freidel:2007py,Engle:2007wy,Kaminski:2009fm}, following quite distinct research logics; the different vertices have only later been recognized as the same. The presentation I have given here does not follow any of the original derivations, and is taken from \cite{Bianchi:2010Nice}.

Quite astonishingly, the simple and natural vertex amplitude \eq{vertex} seems to yield the Einstein equations in the large distance classical limit, as I will argue below. A natural group structure based on $SU(2)\subset SL(2,\C)$  appears to turn out to code the Einstein equations. 

The incredulity called by the surprise for this claim is perhaps tempered by two considerations. The first is that the same happens in QED.  The simple vertex amplitude
\ba
   \langle W| \psi_{e_1}^A(p_1), \psi_{e_2}^B(p_2), \psi_\mu^\gamma(k) \rangle
   &=& \raisebox{-.5cm}{\includegraphics[scale=0.4]{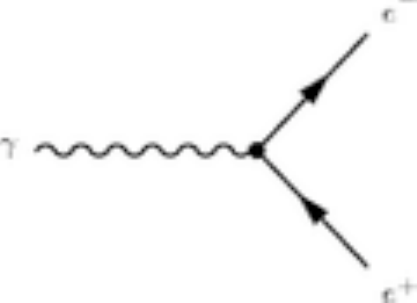}}\\
   &=&
   e\ \gamma_\mu^{AB}\ \delta (p_{1}\!\!+\!p_{2}\!\!+\! k )\nonumber
\ea
yields the full complexity of the interacting Dirac-Maxwell equations. In other words, QED, with its fantastic phenomenology and its 12 decimal digits accurate predictions, is little more that momentum conservation plus the Dirac matrices $\gamma_\mu^{AB}$, which, like $f_\gamma$, are essentially Clebsch-Gordan coefficients.

The second consideration is that general relativity is $BF$ theory plus the simplicity constraints. $BF$ theory means flat curvature. Hence in a sense GR is flat curvature plus simplicity conditions \eq{simp}. The map $f_\gamma$ implements the simplicity conditions, since it maps the states to the space where the simplicity conditions hold (weakly); while the evaluation on $H_l=\id$ codes (local) flatness. \\[-2.mm]

{\footnotesize  The last observation does not imply that the theory describes flat geometries, for the same reason for which Regge calculus describes curved geometries using \emph{flat} 4-simplices. In fact, there is a derivation of the vertex (\ref{vertex}) which is precisely based on Regge calculus, and a single vertex is interpreted as a flat 4-simplex \cite{Engle:2007qf,Engle:2007wy}. 

In this derivation one only considers 4-valent nodes and 5-valent vertices. On the other hand, the resulting expression naturally generalized to an arbitrary number of nodes and vertices, and therefore defines the dynamics in full LQG. This fact was nicely emphasized in \cite{Kaminski:2009fm}.

}\vskip.3cm

\begin{figure}[h]
\centerline{\includegraphics[scale=0.3]{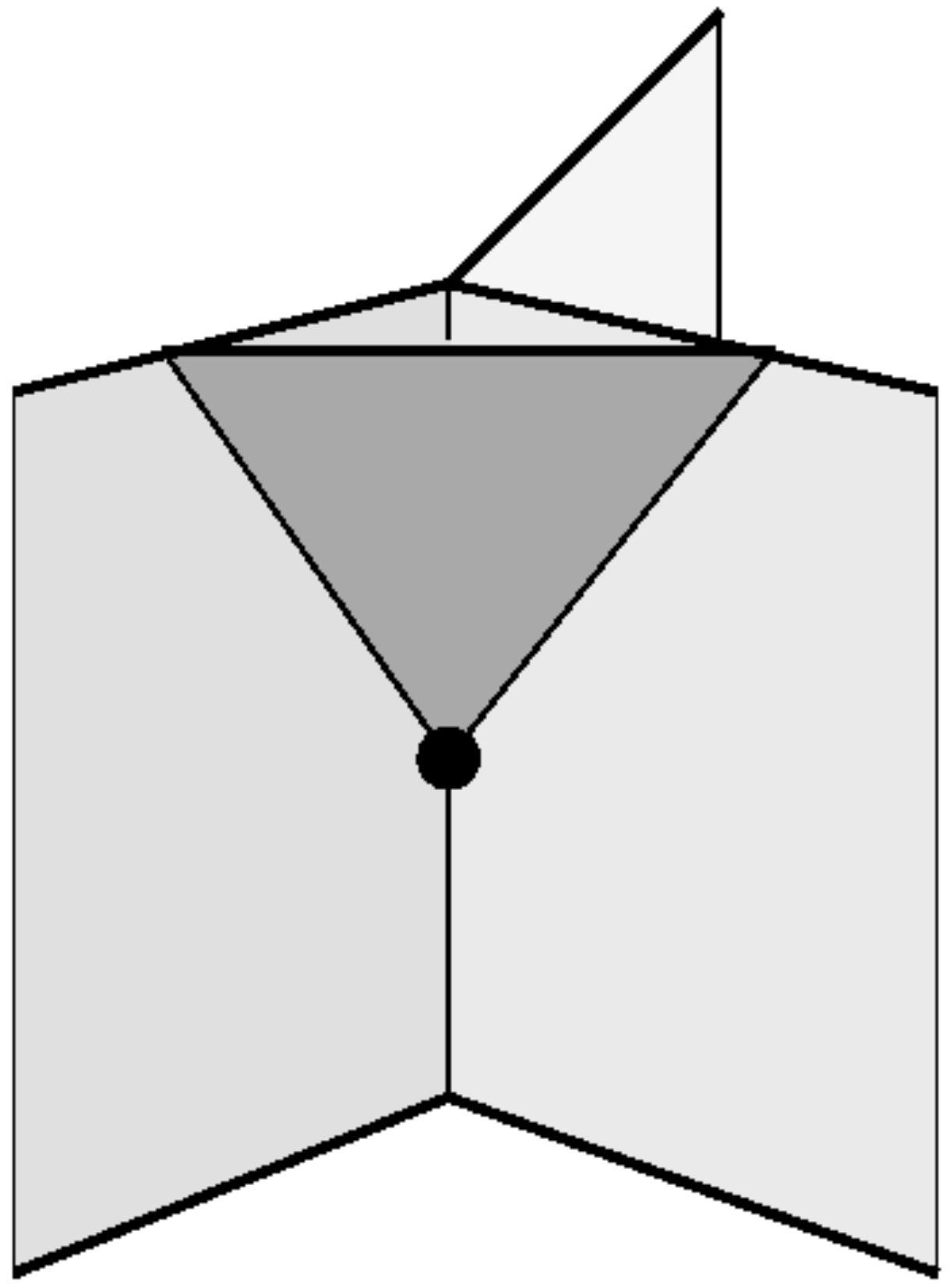}}
\caption{The transition between a single grain of space to three grains of space.}
\label{13}
\end{figure}

The vertex amplitude (\ref{vertex}) gives the probability amplitude for a single spacetime process, where $n$ grains of space are transformed into one another. It has the same {\em crossing} property as standard QFT vertices. That is, it describes different processes, obtained by splitting differently the boundary nodes into ``in" and ``out" ones. For instance if $n=5$ (this is the case corresponding to a 4-simplex in the triangulation picture), the vertex (\ref{vertex})  gives the amplitude for a single grain of space splitting into four grains of space; or for two grains scattering into three, and so on.  A picture of the vertex of Figure \ref{vrtx} interpreted as a 1 to 3 transition, with the future upward,  is in Figure \ref{13}.

More precisely, the vertex $\langle W_v|\psi\rangle$ gives an amplitude associated to the spacetime process defined by a finite region of spacetime, bounded by a 3d region described by the state $\psi$: there is no distinction between ``in" and ``out" states. 

The amplitude (\ref{vertex}) can be written explicitly as \cite{Bianchi:2010mw}
\ba 
W_v(U_{l})&\!\equiv&\! \bk{W_v}{U_l}
\\
&=&\int_{SL(2,\C)^N}  d\tilde g_n \, ~
\prod_{\ell}     P(U_{l} \, , \,g_{s(l)} g_{t(l)}^{-1} )
\label{daniele}\nonumber
\ea
where 
\be
P(U,g)\! =\! \! 
\sum_{j} {\scriptstyle (2j+1)}\, {\rm Tr}\!\!
\left[
D^{\scriptscriptstyle(j)}\!(U)Y_\gamma^\dagger D^{\scriptscriptstyle(\gamma j,j)}(g)Y_\gamma
 \right]\!.
\label{daniele2}
\ee
The integral is over $SL(2,\C)^N$; it implements the projection $P_{SL(2,\C)}$. The measure is $d \tilde g_n =\delta(g_1)dg_n$; the delta function avoids the divergence and does  not spoil gauge invariance  \cite{Engle:2008ev}.

The same amplitude takes a more manageable form when written in terms of coherent states. First, it is easy to show that in terms of LS states, it reads
\be
 W_v(j_l,\vec n_l,\vec n'_l) =\! \int\! d\tilde g_n  \,
 \bigotimes_l  \ 
 \langle \vec n_{l} |g_{s(l)} g_{t(l)}^{\scriptscriptstyle -1}|\vec n'_{l}\rangle_{(\gamma j,j)}
\label{lorentzian}
\ee
The scalar product is taken in the irreducible $SL(2,\C)$ representation $H_{(\gamma j,j)}$ and $|\vec n_l\rangle$ is the coherent state $|j,\vec n_l\rangle$ sitting in the lowest spin subspace of this representation. 

Second, the form of the vertex in the holomorphic basis defined by the coherent states \eq{states} can be obtained combining the definition \eq{vertex} of the vertex and the definition \eq{states} of the coherent states. This gives \cite{Bianchi:2010mw}
 \ba 
W_v(H_{l})&\!\equiv&\! \bk{W_v}{\psi_{H_\ell}}
\\
&=&\int_{SL(2,\C)^N}  d\tilde g_n \, ~
\prod_{\ell}     P(H_{l} \, , \,g_{s(l)} g_{t(l)}^{-1} )
\label{daniele}\nonumber
\ea
where 
\be
P(H,g)\! =\! \! 
\sum_{j} {\scriptstyle (2j+1)}\,e^{\scriptscriptstyle- j(j+1)t}\ {\rm Tr}\!\!
\left[
D^{\scriptscriptstyle(j)}\!(H)Y_\gamma^\dagger D^{\scriptscriptstyle(\gamma j,j)}(g)Y_\gamma
 \right]\!.
\label{daniele2}
\ee
Here $D^{(j)}$ is the analytic continuation of the Wigner matrix from $SU(2)$ to $SL(2,\C)$ and $Y_\gamma$ is defined in \eq{Y}. This is the ``holomorphic" form of the vertex amplitude. 

\subsection{Spinfoams}

By bringing together equations \eq{Wpsi}, \eq{Wsigma} and \eq{vertex} we have the full amplitude associated to a boundary spin network state. 
This can be expressed as the ``spinfoam sum"
\be
\langle W|\psi\rangle=\sum_\sigma \ \prod_f d(j_f)   \prod_v W_v(\sigma).
\label{sf}
\ee
The sum%
\footnote{{\em Note added in proofs:} With a slight modification of the amplitude, the sum can be shown to be equivalent to the limit in which the foam is infinitely refined \cite{Rovelli:2010qx}.}
 is over spinfoams $\sigma$ bounded by the spin network $\psi$. A spinfoam is a two-complex colored with representations on the faces and intertwiners on the links. A two-complex is a collection of faces $f$ meeting at edges $e$, in turn meeting at vertices $v$. The coloring of a face $f$ is a $SU(2)$ representation $j_f$. The coloring of an edges $e$ is given by an element $i_e$ chosen among a basis of vectors in ${\cal H}_{e}={\rm Inv}_{SU(2)}[\bigotimes_{f\in e}  H_{j_f}]$, where the sum is over the faces $f$ bounded by $e$. 

If we ``cut a spinfoam with a 2d-surface", we obtain a spin network: the intersection of the edges $e$ with the surface gives the nodes $n$ of the spin network and the 
intersection of the faces $f$ with the surface gives the links $l$ of the spin network, with their respective colorings.  

In particular, an $S_3$ surface surrounding a vertex $v$ of $\sigma$ defines a spin network $\psi_v$. The vertex amplitude of the vertex $v$ of $\sigma$ is defined to be 
\be
W_v(\sigma):= \langle W_v|\psi_v\rangle.
\ee
This is ``local", in the sense that it depends only on the spins and intertwiners surrounding the vertex.

The amplitude of a spinfoam is the product of the amplitudes $W_v$ of the single vertices, times the product of face amplitudes, needed to obtain the proper inner product when gluing boundary spaces \cite{Bianchi:2010fj}.  This gives the dimension of the $SU(2)$ representation coloring the face: $d(j_f)=(2j_f+1)$.\\[-2.mm]

{\footnotesize  Alternative choices for the face amplitude have been considered in the literature. In the Euclidean case, where $SL(2,\C)$ is replaced by $SO(4)$, there is a natural alternative which is the dimension of the $SO(4)$ irreducible into which the representation $j$ is mapped by $Y_\gamma$.  I suspect that this choice is incompatible with the natural requirement that the Hilbert-space contraction of the amplitudes of two spinfoams along a common boundary be the same as the amplitude of the composed spinfoam.  

A simple modification of the theory is to multiply the vertex by a constant $\lambda$. This comes naturally if one derives \eqref{sf} from a group field theory \cite{Oriti:2009wn}: then $\lambda$ is the coupling constant in front of the group-field-theory interaction term.  The physical interpretation of the constant $\lambda$ is debated \cite{Oriti:2009wn,Ashtekar:2010ve}. 

}\hspace{1em}

The expression (\ref{sf}) fully defines a quantum field theory of gravity.  All that remains to do is to extract physics from this theory, and show that it gives general relativity in some limit. \\[-2.mm]

{\footnotesize    (The last paragraph is an over-statement.)}

\subsection{The euclidean theory}

Before describing how to use the above definition of the dynamics, it is useful to introduce also ``euclidean quantum gravity", which is the theory obtained from the one above by replacing $SL(2,\C)$ with $SO(4)$. The representations of $SO(4)$
are labelled by two spins $(j^+,j^-)$. The theory is the same as above with the only difference that (\ref{p}) is replaced by 
\be
          j^\pm=\frac{|1\pm \gamma|}2
\ee
and $f_\gamma$ maps $H_j$ into the lowest spin component of $H_{j^\pm}$ if $\gamma>1$, but to  the highest spin component of $H_{j^\pm}$ if $\gamma<1$ (the case $\gamma=1$ is ill defined.) All the rest goes through as above. The vertex amplitude can be written in the simpler form
\be
 W_v(j_l,\vec n_l,\vec n'_l) =\! \int\! dg^\pm_n  \,
 \bigotimes_l  \prod_{i=\pm}
 \langle \vec n_{l} |g^i_{s(l)} (g^i_{t(l)})^{\scriptscriptstyle {\rm-}1}|\vec n'_{l}\rangle^{2j^i}
\label{euclidean}
\ee
where now the integration is over $SU(2)^N\times SU(2)^N\sim SO(4)^N$ and the scalar product is in the fundamental representation of $SU(2)$.

\subsection{Transition amplitudes}

The predictions of the theory are in the transition amplitudes. Given a boundary state, the formalism defined above can be used to define transition amplitudes, namely to associate probabilities to boundary states (processes). We are particularly interested in processes involving (background)  semiclassical geometries.  Since the formalism is background independent, the information about the background over which we are computing amplitude must be fed into the calculation. This can only be done with the choice of the boundary state.

Consider a three-dimensional surface $\Sigma$ with the topology of a three sphere. Let $(q,k)$ be the three-metric and the extrinsic curvature of $\Sigma$.    The \emph{classical} Einstein equations determine uniquely whether or not $(q,k)$ are physical: that is, whether or not there exist a Ricci-flat spacetime $\cal M$ (a solution of the Einstein equation) which is bounded by $(\Sigma, q, k)$.\footnote{This is the analog of the following formulation of dynamics. Given coordinate and momenta $q_0,p_0,q_t,p_t$ at $t$=$0$ and at a final time $t$, dynamics is fully captured by the conditions the quadruplet $(q_0,p_0,q_t,p_t)$ must satisfy in order to bound a physical trajectory. For a free particle, for instance, these are $p_t=p_0=m(q_t-q_0)/t.$} The \emph{quantum} theory will assigns an amplitude to any semiclassical boundary state peaked on a given boundary geometry $(q, k)$, and we expect this amplitude to be suppressed if $(\Sigma, q, k)$ does not bound a solution of the Einstein equations. 

Now, a consider a boundary state in the holomorphic representation --- this can be given an interpretation as a classical geometry, as discussed above.  Choose a (normalized) holomorphic coherent state $\psi_{H_l}$ determined by a discrete geometry $H_l$ that approximates $(g,k)$ in a suitable approximation. Then, if $q,k$ is a solution of the Einstein equations, we must expect that, within the given approximation
\be
P(\psi_{H_l})= |\langle W|\psi_{H_l}\rangle|^2 \sim 1.
\ee

Next, if we modify the state $\psi_{H_l}$ with field operators $E_1,... ,E_n$, then the amplitude
\be
W_{H_l}(E_1,...,E_n) =  \langle W|E_1... E_n | \psi_{H_l}\rangle.
\ee
can be interpreted as a scattering amplitude between the $n$ ``particles" (quanta) created by the field operators over the spacetime  $\cal M$. (The possibility of using the notion of ``particle" in this context is discussed in detail in \cite{Colosi:2004vw}.) Since we know how to write the gravitational field operator (the triad), we can in principle compute graviton $n$-point functions in this way.   

\section{Expansions}

There is no physics without approximations. The full sum \eq{sf} is intractable, as far as we can see. It can be compared to the full perturbation expansion in QED.  (But see \cite{Oriti:2009wn}.)  We need some appropriate way to compute approximate transition amplitudes, as we do in for instance order by order in perturbative QED, or on a finite lattice in QCD. 

What approximations can be effective in the background-independent context of quantum gravity? I consider here three expansions that naturally present themselves.

\subsubsection{Graph expansion} 

Consider the component ${\cal H}_\Gamma$ of ${\cal H}$.  Notice that because of the equivalence relation defined in Section \ref{Hs}, all the states that have support on graphs smaller than (subgraphs of) $\Gamma$ are already contained in  ${\cal H}_\Gamma$, provided that we include also the $j=0$ representations. Therefore if we truncate the theory to a single Hilbert space  ${\cal H}_\Gamma$ for a given fixed $\Gamma$, what we loose are only states that need a ``larger" graphs to be defined. Let us therefore consider the truncation of the theory to a given graph.\footnote{The analog in QFT is to truncate the theory to the sector of Fock space with a 
number of particles less than a finite fixed maximum number.  Notice that virtually {\em all} calculations in perturbative QED are performed within this truncation.}

What kind of truncation is this?  It is a truncation of the degrees of freedom of general relativity down to a finite number; which can be interpreted as describing the lowest modes on a mode expansion of the gravitational field on a compact space. Strictly speaking this is neither an ultraviolet nor an infrared truncation, because the whole space can still be large or small. What are lost are not wavelengths shorter that a given length, but rather wavelengths  $k$ times shorter the full size of space, for some integer $k$.  

It is reasonable to expect this truncation to define a viable approximation for all gravitational phenomena such that the ratio between the largest and the smallest relevant wavelengths in the boundary state is not large. For instance, conventional cosmology is based on a truncation of general relativity to a single degree of freedom, the scale factor. Similarly, the scattering amplitude of modes with wavelength $\lambda$ is dominated by the physics of the degrees of freedom with wavelength of the order of $\lambda$. And so on.  The approximation can then be improved by taking a larger graph. 

Notice that fixing the boundary graph does not mean that we are taking the approximation in which the dynamics does not change the graph. The bulk two complex can still be arbitrary.

Finally, notice that the graph expansion resolves the problem given by the fact that the operators of the theory are defined on ${\cal H}_\Gamma$ rather than on $\cal H$.  

\subsubsection{The vertex expansion}

A natural expansion of (\ref{sf}) presents itself: the expansion in the number $N$ of vertices of $\sigma$.  

In which regime is this expansion useful?  We have a hint of this from the Regge interpretation of the vertex amplitude: if we derive the vertex amplitude from a Regge discretization of general relativity, a single vertex corresponds to a {\em flat} 4-simplex.  It is therefore natural to expect that cutting the theory to small $N$ defines an approximation valid around flat space, and where relevant wavelengths are not much shorter than the bounded scattering region $\cal R$. 

\begin{figure}[h]
\centerline{\includegraphics[scale=0.03]{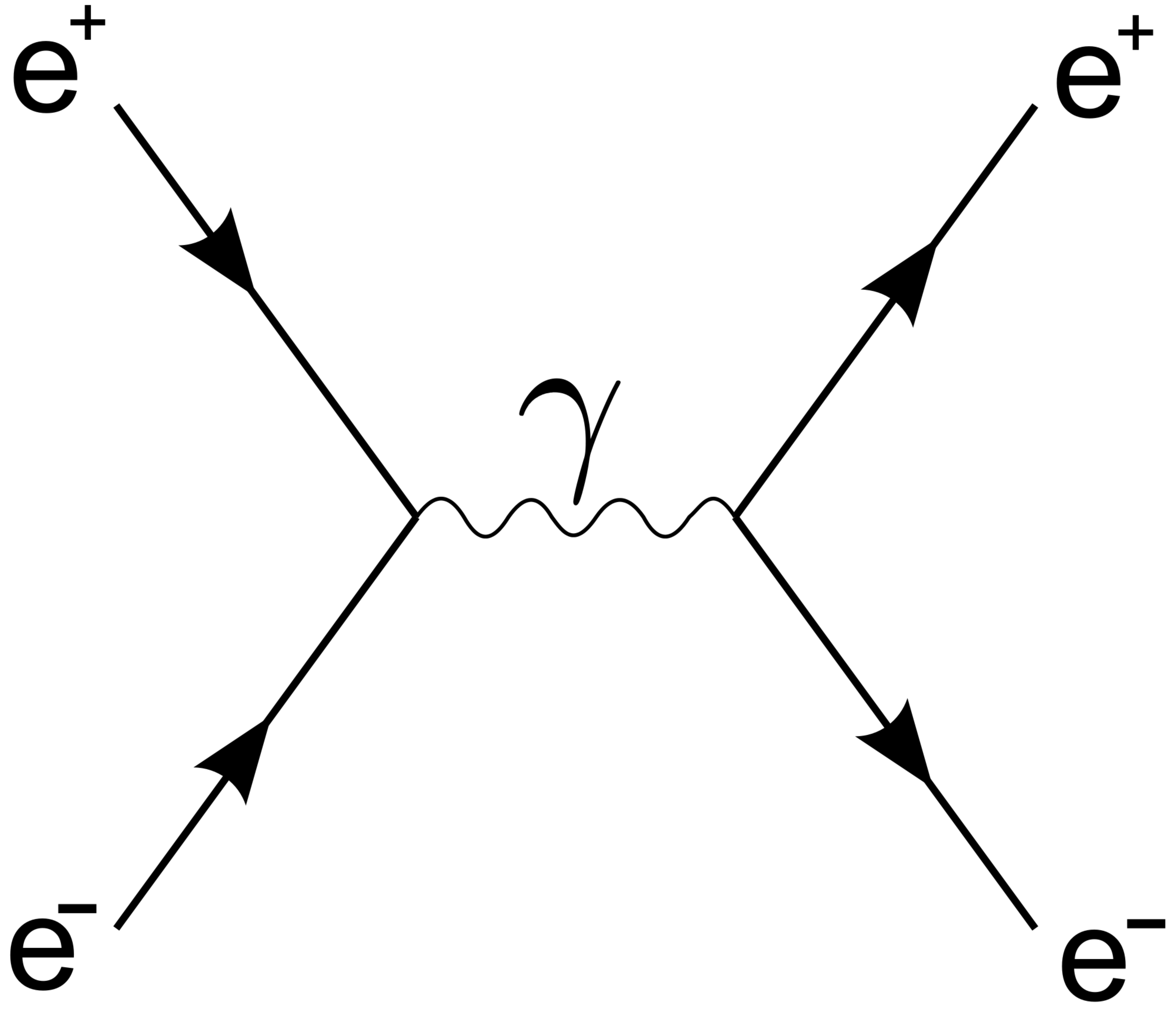}}
\caption{Electromagnetic scattering of two electrons at first relevant order in QED.}
\label{ee}
\end{figure}

Notice the similarity of this expansion with the standard perturbation expansion of QED. In both cases, we describe a quantum field in terms of interactions of a finite number of its ``quanta".  In the case of QED, these are the photons.   In the case of LQG, these are the ``quanta of space", or ``chunks of space", described in Section \ref{interpr}. In the QED case, individual photons can have small or large energy; in the quantum gravity case, the quanta of space can have small or large volume.  In the case of QED, one should be careful not to take the photon picture too literally when looking a the semiclassical limit of the theory. For instance, the Feynman graph for the Coulomb   scattering of two electrons is given in Figure \ref{ee}. But Figure \ref{ee} does \emph{not} provide a viable picture of the continuous electric field in the scattering region.   Similarly, if we compute a transition amplitude between geometries at first order in the vertex expansion, we should not mistake the corresponding spinfoam for a faithful geometrical picture of the gravitational field in the corresponding classical spacetime. 

An important observation regards radiative corrections. The QED perturbative expansion is viable because the effect of all the radiative corrections due to the higher frequency modes can be absorbed into the renormalization of a few parameters.  Does the same happen in LQG?  For the moment, this is not known. Preliminary calculations are encouraging: they indicate \emph{finite} radiative corrections of the vertex and \emph{logarithmic} correction for the ``self energy" \cite{Perini:2008pd}\footnote{When \cite{Perini:2008pd} was written, the choice of the face amplitude was still very unclear. I find remarkable that later independent arguments point towards the most favorable choice ($d_j\sim j$).}, but these results are preliminary.  

Potential divergences in the theory all infrared, not ultraviolet, because there is no 
short-scale geometry (sub Planckian geometry) in the theory.

\subsubsection{The large distance expansion}

Finally, a useful approximation can be taken by choosing the boundary to be large.  This means that the boundary state must be peaked on a boundary geometry which is large compared with the Planck length. In particular, we can chose holomorphic  boundary states $\psi_{H_l}$ where $\eta_l\gg1$ in each $H_l$. 

The analysis of the vertex (\ref{lorentzian}) as well as that of its euclidean analog (\ref{euclidean}) in this limit has been carried out in great detail for the 5-valent vertex, by the Nottingham group \cite{Barrett:2009mw,Barrett:2009gg,Barrett:2009cj,Pereira:2010}.  The remarkable result of this analysis is that in this limit the vertex behaves as 
\be
W_v \sim
  e^{i S_{Regge}}
\label{not}
\ee
where $S_{Regge}$ is a function of the boundary variables given by the Regge action, under the identifications of these with variables describing a Regge geometry. The Regge action codes the Einstein equations' dynamics.  Therefore this is an indication that the vertex can yield general relativity in the large distance limit. More correctly, this result supports the expectation that the boundary amplitude reduces to the exponential of the Hamilton function of the classical theory. \\[-2.mm]

{\footnotesize In fact, what is shown in \cite{Barrett:2009mw} is that $W_v\sim e^{i S_{Regge}}+ e^{-i S_{Regge}}$.  Concern has been raised by the fact that two terms appear in this sum. In my opinion this concern is excessive. When sandwiched between coherent  boundary states that define a \emph{semiclassical} geometry, only one of the terms in survives \cite{Bianchi:2010mw}. This is because of the ubiquitous mechanism of phases cancellations between propagator and boundary state in quantum mechanics. See \cite{Bianchi:2006uf} for a discussion of this mechanism. Therefore the existence of different terms in (\ref{not}) does not affect the classical limit of the theory.

On the other hand, I think that the amplitude of the theory \emph{should} include different terms. This appears clearly in the three dimensional Ponzano Regge theory \cite{Rovelli:1993kc} as well as in low dimensional models \cite{Colosi:2003si}, and can be viewed as related to the fact that the classical dynamics does not distinguish propagation ``ahead in (proper) time" or ``backward in (proper) time", in a theory where coordinate time is an unphysical parameter.\footnote{It is sometime argued that the presence of the two terms follows from the fact that one has failed to select the ``positive energy" solutions in the course of the quantization.  However, such choice makes only sense in the context of the specific strategy for quantization which consists in considering \emph{complex} solutions of the classical equations and then \emph{discarding} solutions with ``negative energy". This strategy is not available here, because of the absence of a preferred time, or a preferred energy. But there are other quantization strategies that are available: we quantify the \emph{real} solution space and keep \emph{all} solutions.  In other words, the physical scalar product is determined by all \emph{real} solutions of the Wheeler DeWitt equation with the proper symplectic structure, not by a ``positive energy sector" of the complex solutions.}

}

\subsection{What has already been computed}

Using the approximations discussed above, a few transition amplitudes have already been computed in the (Euclidean) theory. 

\subsubsection{$n$-point functions}

The two point function of general relativity over a flat spacetime has been computed by Bianchi, Magliaro and Perini in  \cite{Bianchi:2009ri}, following the earlier attempts in \cite{Rovelli:2005yj,Bianchi:2006uf,Alesci:2007tx,Alesci:2007tg,Alesci:2008ff} and using the Euclidean theory rather that the Lorentzian one (that is, using (\ref{euclidean}) instead than (\ref{lorentzian})), and has been shown to converge to the free graviton propagator of quantum gravity in the large distance limit. 

The calculation has been performed to first order in the vertex expansion, on the complete graph with five nodes $\Gamma_5$, and to first order in the large-distance expansion. The boundary state $\psi_L$ has been chosen as the coherent state determined by the (intrinsic and extrinsic) geometry of the boundary of a \emph{regular} 4-simplex\footnote{This approximates flat space.  More precisely, the quantity computed can be interpreted as the graviton two-point function \emph{under the condition} that the state of the gravitational field is described by this spin network at large wavelength. In other words, what is assumed is not the full state of the gravitational field, but only the value of some of its variables. Intuitively, this can be seen as the translation into the theory of a \emph{finite} number of macroscopic geometrical measurements that measure flat space.} of size $L$. 
\begin{equation}
\Gamma_5 = \hspace{-5em}\setlength{\unitlength}{0.0004in} 
\begin{picture}(5198,1100)(5000,-4330)
\thicklines
\put(8101,-5161){\circle*{68}}
\put(8401,-3961){\circle*{68}}
\put(6601,-3961){\circle*{68}}
\put(7501,-3361){\circle*{68}}
\put(6901,-5161){\circle*{68}}
\put(8101,-5161){\line(1,4){300}}
\put(7501,-3361){\line( 3,-2){900}}
\put(6601,-3961){\line( 3, 2){900}}
\put(8101,-5161){\line(-1, 0){1200}}
\put(6901,-5161){\line(-1, 4){300}}
\put(7876,-4479){\line( 1,-3){225.800}}
\put(7089,-4351){\line(-6, 5){491.312}}
\put(7801,-3961){\line( 1, 0){600}}
\put(6901,-5161){\line( 1, 3){383.100}}
\put(7321,-3871){\line( 1, 3){173.100}}
\put(7501,-3354){\line( 1,-3){325.500}}
\put(7456,-4726){\line(-4,-3){581.920}}
\put(8394,-3961){\line(-5,-4){828.902}}
\put(7569,-4629){\line( 0,-1){  7}}
\put(6601,-3969){\line( 1, 0){1020}}
\put(8109,-5161){\line(-5, 4){867.073}}
\end{picture}\hspace{-3em}.
\label{fsimpic}
\end{equation}
\vskip1cm
The quantity computed is 
\ba
W^{abcd}_{mn} &=&  \langle W| \vec L_{na}\cdot \vec L_{nb}\  \vec L_{mc} \cdot \vec L_{md} | \psi_L\rangle \label{prop}
\\
&& -\langle W| \vec L_{na}\cdot \vec L_{nb} | \psi_L\rangle\langle W\vec L_{mc} \cdot \vec L_{md}^j | \psi_L\rangle.\nonumber
\ea
where $m,n,a,b...=1,...,5$ label the nodes of $\Gamma_5$. The resulting expression can be compared with the corresponding quantity 
\ba
W^{abcd}(x_m,x_n) &=&  \langle 0| g^{ab}(x_n)g^{cd}(x_m) | 0\rangle \\
&& -  \langle 0| g^{ab}(x_n)| 0\rangle \langle 0|g^{cd}(x_m) | 0\rangle.\nonumber
\ea
in conventional QFT, where $g^{ab}(x)$ is the gravitational field operator.

In this way, it is clear that $n$ point functions in gravity can be computed order by order.

\subsubsection{Cosmology}

The transition amplitude between two homogeneous and isotropic coherent states in quantum cosmology has been computed in \cite{Bianchi:2010zs}. The calculation is: (i) in the approximation where the theory is truncated on the graph formed by two copes of the graph $\Delta_2^*$;  $\Delta_2^*$ is the ``dipole" graph formed by two nodes connected by four links \cite{Rovelli:2008dx}
\begin{center}
\begin{picture}(20,40)
\put(-36,28) {$\Delta_2^*\  = $}
\put(04,30) {\circle*{3}} 
\put(36,30) {\circle*{3}}  
\qbezier(4,30)(20,53)(36,30)
\qbezier(4,30)(20,21)(36,30)
\qbezier(4,30)(20,39)(36,30)
\qbezier(4,30)(20,7)(36,30)
\put(50,27) {\circle*{1}} 
\end{picture}
\end{center} 
\vspace{-1.5em}%
and (ii) at first order in the large distance expansion and in the vertex expansion. The spinfoam considered has therefore the form
\begin{center}
\begin{picture}(20,30)\setlength{\unitlength}{0.012in} 
\put(04,30) {\circle*{3}} 
\put(36,30) {\circle*{3}}  
\qbezier(4,30)(20,53)(36,30)
\qbezier(4,30)(20,21)(36,30)
\qbezier(4,30)(20,39)(36,30)
\qbezier(4,30)(20,7)(36,30)
\put(04,-30) {\circle*{3}} 
\put(36,-30) {\circle*{3}}  
\qbezier(4,-30)(20,-53)(36,-30)
\qbezier(4,-30)(20,-21)(36,-30)
\qbezier(4,-30)(20,-39)(36,-30)
\qbezier(4,-30)(20,-07)(36,-30)
\linethickness{0.4mm}
\put(20,0) {\circle*{6}} 
\qbezier(4,29)(20,0)(36,-29)
\qbezier(4,-29)(20,0)(36,29)
\end{picture} 
\vspace{1cm}
\end{center} 
Homogenous isotropic states depend on two variables, $p$ and $c$, at each $\Delta_2^*$. These represent (the square of) the radius $a$ and the extrinsic curvature of closed universe (or $\dot a$).  They enter the definition of the holomorphic coherent states via a complex combination $z\sim c+ip$. The resulting transition amplitude turns out to be 
\be
 W(z,z') \sim zz' e^{-\frac{z^2+{z'}^2}{2t\hbar}}.
\ee
This amplitude reproduces the correct Friedmann dynamics in the sense that it satisfies a quantum constraint equation which reduces to the (appropriate limit of the) Friedmann hamiltonian in the classical limit \cite{Bianchi:2010zs}.

\section{Open problems}\label{problems}

The theory is far from being complete. Here are some of the open problems that require further investigation. 
\begin{enumerate}
\item Compute the propagator (\ref{prop}) in the Lorentzian theory, extending the euclidean result of  \cite{Bianchi:2009ri}.
\item Compute the three point function and compare it with the vertex amplitude of conventional perturbative quantum gravity on Minkowski space.
\item Compute the next vertex order of the two point function, for $N=2$.
\item Compute the next graph order of the two point function, for $\Gamma>\Gamma_5$.
\item Understand the normalization factors in these terms, and their relative weight. Find out under which conditions the expansion is viable.
\item Study the radiative corrections in (\ref{sf}) and their possible (infrared) divergences, following the preliminary investigations in \cite{Perini:2008pd}. In particular, the sum can be split into a sum over two complexes and a sum over labelings (spin and intertwiners) for a given two complex. The potential divergences of the second are associated to ``bubbles" (nontrivial elements of the second homotopy class) in the two complex. Classify them and study how do deal with these. 
\item Use the analysis of the these radiative corrections to study the scaling of the theory. 
\item In particular, how does $G$ scale?
\item Study the quantum corrections that this theory adds to the tree-level $n$-point functions of classical general relativity. Can any of these be connected to potentially observable phenomena?
\item Is there any reason for a breaking or a deformation of local Lorentz invariance, that could lead to observable phenomena such as $\gamma$ ray bursts energy-dependent time of arrival delays, in this theory?
\item Compute the cosmological transition amplitude in the Lorentzian theory, extending the euclidean result of  \cite{Bianchi:2010zs}. Compare with canonical Loop Quantum Cosmology \cite{Ashtekar:2008zu,Bojowald:2006da}. 
\item The possibility of introducing a spinfoam-like expansion starting from Loop Quantum Cosmology has been considered by Ashtekar, Campiglia and Henderson  \cite{Ashtekar:2009dn,Ashtekar:2010ve,Rovelli:2009tp,Henderson:2010qd}. Can the convergence between the two approaches be completed? 
\item Find a simple group field theory \cite{Oriti:2009wn} whose expansion gives (\ref{sf}).
\item Find the relation between this formalism and the way dynamics can be treated in the canonical theory.  Formally, if $H$ is the Hamiltonian constraint, we expect something like the main equation
\be
               HW=0
\ee
or $WP=0$ where the operator $P$ is given by $\langle W|\overline\psi\otimes\phi\rangle=\langle \psi | P|\phi\rangle$, since $P$ is formally a projector on the solutions of the Wheeler de Witt equation 
\be
H\psi =0.
\ee
Can we construct the Hamiltonian operator in canonical LQG such that this is realized? 
\item Is the node expansion related to the amount of boundary data available? How? 
\item Where is the cosmological constant in the theory? It is tempting to simply replace \eq{vertex} with a corresponding quantum group expression 
\be
   \langle W_v| \psi\rangle =   Ev_q(f\psi).
\label{vertex2}
\ee
where $Ev_q$ is the quantum evaluation in $SL(2,\C)_q$. Does this give a viable theory? Does this give a finite theory?  

\item How to couple fermions and YM fields to this formulation?  The kinematics described above generalizes very easily to include fermions (at the nodes) and Yang Mills fields (on the links). Can we use the simple group theoretical argument that
has selected the gravitational vertex also for coupling these matter fields? 
\end{enumerate}

In conclusion, the theory looks simple and beautiful to me, both in its kinematical and its dynamical parts.  Some preliminary physical calculations have been performed and the results are encouraging. The theory is moving ahead fast.  But we do not yet know if it really works, and there is still very much to do. 


\end{document}